\documentclass[pra,aps,twocolumn,showpacs,superscriptaddress,amsmath,amssymb]{revtex4}
\usepackage{graphicx}
\usepackage{graphics}
\usepackage{epsfig}
\usepackage{dcolumn}
\usepackage{bm}
\begin{document}
\title{Preparation of atomic Fock states by trap reduction}

\author{M. Pons}
\email{marisa.pons@ehu.es}
\affiliation{Departamento de F\'isica Aplicada I,
E.U.I.T. de Minas y Obras P\'ublicas, Universidad del Pa\'\i s Vasco, 48901 Barakaldo, Spain}

\author{A. del Campo}
\email{a.del-campo@imperial.ac.uk}
\affiliation{Institute for Mathematical Sciences, Imperial College London, SW7 2PE, UK;\\
QOLS, The Blackett Laboratory, Imperial College London, Prince Consort Rd., SW7 2BW,UK}

\author{J. G. Muga}
\email{jg.muga@ehu.es} 
\affiliation{Departamento de Qu\'\i mica-F\'\i sica, Universidad del Pa\'\i s Vasco, Apartado 644, 48080 Bilbao, Spain}

\author{M. G. Raizen}
\email{raizen@physics.utexas.edu}
\affiliation{Center for Nonlinear Dynamics and Department of Physics, University of Texas, Austin, Texas 78712 USA}

\def\d{{\rm d}}
\def\la{\langle}
\def\ra{\rangle}
\def\om{\omega}
\def\Om{\Omega}
\def\vep{\varepsilon}
\def\wh{\widehat}
\def\tr{\rm{Tr}}
\def\da{\dagger}
\newcommand{\beq}{\begin{equation}}
\newcommand{\eeq}{\end{equation}}
\newcommand{\beqa}{\begin{eqnarray}}
\newcommand{\eeqa}{\end{eqnarray}}
\newcommand{\intf}{\int_{-\infty}^\infty}
\newcommand{\into}{\int_0^\infty}
\date{\today}
\begin{abstract}
We describe the preparation of atom-number states with strongly interacting bosons in one dimension, or spin-polarized fermions. The procedure is based on a combination of weakening and squeezing of the trapping potential. For the resulting state, the full atom number distribution is obtained.
Starting with an unknown number of particles $N_i$, we optimize the sudden change in the trapping potential which leads to the Fock state of $N_f$ particles in the final trap. 
Non-zero temperature effects as well as different smooth trapping potentials are analyzed. A simple criterion is provided to ensure the robust preparation of the Fock state for physically realistic traps.

\end{abstract}
\maketitle
\section{Introduction}

Given the importance of photon statistics in quantum optics, the field of 
atom statistics is expected to develop vigorously in atom optics, 
fueled by the current ability to measure the number of trapped ultracold atoms with nearly single-atom resolution and without ensemble averaging of fluctuations  \cite{CSMHPR05,Dotsenko05,Schlosser02} . Among the possible atomic distributions, pure    
atom number (Fock) states form a fundamental basis and 
hold unique and simple properties that make them ideal for studies of quantum 
dynamics of few-body interacting systems \cite{Phillips}, precision measurements
\cite{Kasevich}, or quantum information processing
 \cite{ Zoller, Lewenstein,Meystre}. 
Efficient and robust creation, detection and manipulation of atom number states are thus important goals in atomic physics. 
Several approaches have been  
proposed and explored recently with theoretical and experimental work 
leading to sub-Poissonian and, in the limit, number states,
such as atomic tweezers \cite{tweezers1,tweezers2}, 
interferometric methods \cite{interf1,Phillips}, 
Mott insulator states \cite{Mott}, 
or atomic culling \cite{CSMHPR05,DRN07}. None of these methods is so far fully satisfying 
if the individual atoms have to be addressed (a problem of the Mott insulator states
in optical lattices), and if an arbitrary number of atoms is to be produced reliably and with small enough variance for the trapped atom number, 
so further research is still required. 

In a previous paper \cite{DM08} a method was proposed in which the trapping potential is simultaneously weakened and squeezed so that the final trap holds a desired 
number state. For a Tonks-Girardeau gas, it was shown that this mixed trap reduction yields optimal results, even when the process is sudden. From the expression of the number variance, a simple criterion for optimal performance was obtained, namely, that the  subspace spanned by the occupied levels in the initial trap configuration contains the subspace of the bound levels in the final trap. Starting from an unknown number of particles trapped at zero temperature, the mixed trap reduction method assures that the final state indeed corresponds to the desired Fock state by avoiding the momentum or position space truncations inherent in pure squeezing or pure weakening (the latter being called ``culling'' in \cite{DRN07}).           

In \cite{DRN07,DM08} the potential traps considered for simplicity were 
finite square wells, so doubt could be cast on the validity of the
results in actual smooth traps. Other limitations were the consideration of zero temperature initial states, and a
statistical analysis limited to the first and second moment of the number distribution.  
In this paper we overcome these shortcomings by studying the mixed trap reduction process using smooth potentials, states with finite temperature, and  the full number distribution.

\section{The Tonks regime and polarized fermions} 
The strongly interacting regime of ultracold bosonic atoms can be described by the so-called Tonks-Girardeau (TG) gas \cite{Girardeau60}, 
which is achieved  at low densities and/or large one-dimensional scattering length \cite{Olshanii98,PSW00}. It has been argued \cite{DM08}
that this regime is optimal for the creation of atomic Fock states by 
mixed trap reduction. 

The TG gas and its ``dual'' system of spin-polarized ideal fermions behave similarly,
and share the same one-particle spatial density as well as any other local-correlation function, while differ on the non-local correlations. 

The fermionic many-body ground state wavefunction of the dual system is built at time $t=0$ as a Slater determinant for $N_i$ 
particles,
$ \psi_{F}(x_{1},\dots,x_{N_i}) =\frac{1}{\sqrt{N_i!}}{\rm det}_{n,k=1}^{N_i}\varphi_{n}^i(x_{k})$,
where $\varphi_{n}^i(x)$ is the $n-$th eigenstate of the initial trap, whose time evolution will be denoted by $\varphi_{n}(x,t)$ when the external trap
is modified. 
The bosonic wave function, 
symmetric under permutation of particles, is obtained from $\psi_F$ by
the Fermi-Bose (FB) mapping \cite{Girardeau60,CS99} 
$
\psi(x_{1},\dots,x_{N_i})= \mathcal{A}(x_{1},\dots,x_{N_i})\psi_{F}(x_{1},\dots,x_{N_i})
$, 
where $\mathcal{A}=\prod_{1\leq j<k\leq N_i}{\rm sgn}(x_{k}-x_{j})$ is the ``antisymmetric unit function''. Noting that $|\psi|^2=|\psi_{F}|^2$ it is clear that both systems obey the same counting statistics. Moreover, since $\mathcal{A}$  
does not include time explicitly, the mapping is also valid when
the trap Hamiltonian is modified, and the time-dependent density profile resulting from 
this change can be calculated as \cite{GW00b}
%
$\rho(x,t)= N_i\!\!\int\vert\psi(x,x_{2},\dots,x_{N_i};t)\vert^{2} \d x_{2} \cdots\d x_{N_i}
=\sum_{n=1}^{N_i}\vert\varphi_{n}(x,t)\vert^{2}. \label{11}$
%
By reducing the trap capacity (maximum number of bound states and thus particles that it can hold in the TG regime) some of the $N_i$ atoms initially confined
may escape and only $N$ will remain trapped.  

To determine whether or not sub-Poissonian statistics or a Fock state are achieved 
in the reduced trap we need to calculate  
the atom-number fluctuations.

\section{The sudden approximation: full counting statistics}

We shall now describe   
the preparation of Fock states by an abrupt change of the trap potential
to reduce its capacity. 
Consider a trap with an unknown number of particles $N_i$, 
which supports a maximum of $C_i$ bound states.
Generally $N_i$ is smaller than the capacity of the trap $C_i$.
The trapping potential is abruptly modified 
to a final configuration of smaller capacity $C_f$.  
Similarly the final number of trapped particles will be $N_f\le C_f$.
We are interested in the optimal potential change such that $N_f=C_f$ 
to prepare the atomic Fock state $|N_f=C_f\ra$.

Let $\alpha=i,f$ stand for initial and final configuration. 
The Hilbert space associated with the Hamiltonian of a particle moving 
in any realistic trap $V_{\alpha}$, is the direct sum $\mathcal{H}_{\alpha}=\mathcal{B}_{\alpha}\oplus\mathcal{R}_{\alpha}$ 
of the subspace spanned by the bound states $\mathcal{B}_{\alpha}=\{|\varphi_{j}^{\alpha}\ra |j=1,\dots,C_{\alpha}\}$, and that of scattering states $\mathcal{R}_{\alpha}=\{|\chi_{k}^{\alpha}\ra |k\in\mathbb{R}\}$.
Consider the projector onto the
final bound states, $\mathcal{B}_f$ defined as
\beqa
\widehat{\Lambda}_f=\sum_{j=1}^{C_f}|\varphi_j^f\ra\la\varphi_j^f|.
\eeqa 

Within the TG regime and for spin-polarized fermions, the  asymptotic mean number and variance of trapped
atoms are \cite{DM08}
\beqa
\label{mean}
\la N_f\ra&=& \tr(\widehat{\Lambda}_i\widehat{\Lambda}_f)
\eeqa
and  
\beqa
\label{variance}
\sigma_{N_f}^2
&=&\tr\big[\widehat{\Lambda}_i\widehat{\Lambda}_f-(\widehat{\Lambda}_i\widehat{\Lambda}_f)^2\big],
\eeqa
where
\beqa
\widehat{\Lambda}_i=\sum_{n=1}^{N_i}|\varphi_n^i\ra\la\varphi_n^i|
\eeqa 
is the projector onto the bound subspace occupied by the initial state. 
We may thus conclude that trap reduction  
can actually lead to the creation of Fock states 
with $\la N_f\ra=C_f$ and $\sigma^2_{N_f}=0$ quite simply 
when the initial states span the final ones, 
\beqa
\label{cond}
\widehat{\Lambda}_f\subset\widehat{\Lambda}_i.
\eeqa
In fact the full atom number distribution \cite {Levitov} is accessible in the atom culling experiments
\cite{CSMHPR05} and we next focus our attention on it.
Consider the characteristic function of the number of particles in the bound subspace of the final trap,
\beqa
F(\theta)=\tr[\hat{\rho}e^{i\theta\widehat{\Lambda}_f\hat{n}\widehat{\Lambda}_f}].
\eeqa
Following \cite{KKS00}, the atom number distribution can be obtained as its Fourier transform,
\beqa
p(n)=\frac{1}{2\pi}\int_{-\pi}^{\pi}e^{-in\theta}F(\theta)d\theta, 
\eeqa 
with $n=1,\dots,C_f$.

The characteristic function of spin-polarized fermions or a Tonks-Girardeau gas restricted 
to a given subspace was studied in \cite{Klich03,BM04,Schonhammer07}. 
Using the projector for the bound subspace in the final configuration,
\beqa
F(\theta)={\rm det}[1+(e^{i\theta}-1)\widehat{\Lambda}_f\widehat{\Lambda}_i].
\eeqa
For computational purposes it is convenient to use the basis of single-particle eigenstates $|\varphi_m^f\ra$ which spans 
the final bound subspace so that $F(\theta)={\rm det}{\bf A}$, where ${\bf A}$ is a $C_f\times C_f$ matrix with elements 
\beqa A_{nm}=\delta_{nm}+[\exp(i\theta)-1]\la\varphi_n^f|\widehat{\Lambda}_i|\varphi_m^f\ra.\eeqa
Clearly, if $\widehat{\Lambda}_f\subset\widehat{\Lambda}_i$, $A_{nm}=\exp(i\theta)\delta_{nm}$, $F(\theta)=\exp(i\theta C_f)$ and the Kronecker-delta atom-number distribution associated with the Fock state $|N=C_f\ra$ is obtained, 
\beqa
p(n)=\delta_{n,C_f}.
\label{Fockdistribution} 
\eeqa
For completeness we note that the cumulant-generating function $\log F(\theta)$ admits the expansion
\beqa
\log F(\theta)=\sum_{n=1}^{\infty}\kappa_n\frac{(i\theta)^n}{n!}, 
\eeqa 
from which the mean $\kappa_1=\la N_f\ra$ in Eq. (\ref{mean}) 
 and variance $\kappa_2=\sigma_{N_f}^2$ in Eq. (\ref{variance}) are just the first two orders.

Different regimes of interactions for ultracold Bose gases in tight waveguides can be characterized by a single parameter $\gamma =mg_{1D}L/\hbar^2N$, where $g_{1D}$ is the one-dimensional coupling strength, $L$ the size of the system, $m$ and $N$ the mass and number of atoms respectively.   
$\gamma$ can be varied \cite{Olshanii98} allowing to explore the physics from the mean-field regime ($\gamma\ll1$) to the TG regime ($\gamma\gg 1$) \cite{PSW00}.
We note that for the system to remain in the TG regime, it suffices to keep or decrease the density, since
\beqa
\gamma_f=\gamma_i\frac{n_i}{n_f}.
\eeqa
In particular, trap weakening clearly leads to a reduction of the density so that the system 
goes deeper into the TG regime.

\section{Dependence on the trapping potential}

\begin{figure}[t]
\includegraphics[width=8cm,angle=0]{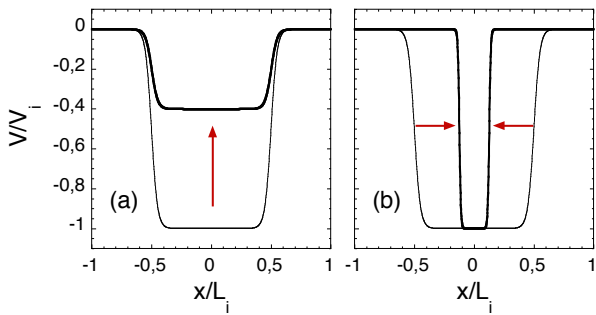}
\caption{\label{potential}
Schematic potential change for trap weakening (a) and squeezing (b), 
relative to the initial width  $L_i$ and depth $V_i$, for $\sigma=0.05L$.
}
\end{figure}

In this section we shall discuss the efficiency of the trap-reduction procedure at zero-temperature, focusing on the relevance 
of the shape of the confining potential.
In particular, instead of the idealized square potentials used in \cite{DRN07,DM08}  we 
shall study here the family of ``bathtub'' potentials
\beqa
\label{trap}
\mathcal{V}_{\alpha}(x;V_{\alpha},L_{\alpha},\sigma_{\alpha})
=-\frac{1}{2}V_{\alpha}
\left[1-\tanh\left(\frac{|x|-L_{\alpha}}{\sigma_{\alpha}}\right)\right]
\eeqa
as well as the inverted Gaussian potential 
\beqa
{\cal V}(x)=-V_\alpha e^{-\frac{x^2}{2\delta^2}}. 
\eeqa
For the bathtub  $L_{\alpha}$ and $V_{\alpha}$ play respectively
the role of the width
and depth of the trap, 
while $\sigma_{\alpha}$ is an additional parameter describing the smoothness
of the potential trap. 

The spectrum and eigenfunctions can be found numerically by a standard technique, first differencing  the Hamiltonian and then diagonalizing the tridiagonal matrix obtained by such difference scheme \cite{Recipes}.

For the bathtub potential, a given $U_{\alpha}=2mV_{\alpha}L_{\alpha}/\hbar^2$ and $\tilde{\sigma}_{\alpha}=\sigma_{\alpha}/L_{\alpha}$ defines a family of isospectral potentials. In dimensionless units, 
their eigenvalues are the same and so are their eigenfunctions. In the limit of a square potential  $U_{\alpha}\approx {C_\alpha}^2\pi^2$. 
For the Gaussian potential a single parameter 
$U_{\alpha}=V_{\alpha}\delta_{\alpha}^2$ defines an isospectral family.

%
\begin{figure}[t]
\includegraphics[width=8.5cm,angle=0]{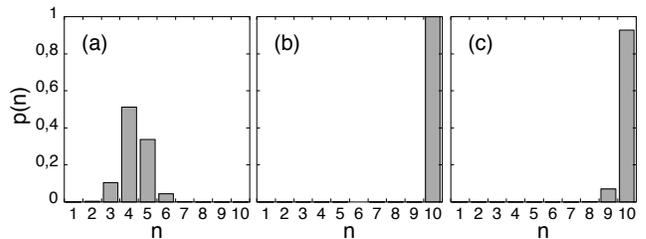}
\caption{\label{figFockdistribution}
Atom number distribution $p(n)$ in a trap reduction setup combining both squeezing and weakening techniques, for a bathtub potential given by Eq.(\ref{trap})  with  $\sigma_\alpha=0.03L_\alpha$ and the parameters of the trap being $N_i=C_i=100$ and  $C_f=10$ ($U_i=10^4\pi^2$, $U_f=10^2\pi^2$). Plot (a) corresponds to almost pure squeezing, with  $L_f/L_i=0.04$, while plot (c) corresponds to pure weakening,  $L_f/L_i=1,$ showing that it is the combination of both techniques, $L_f/L_i=0.5$, plot (b), the most efficient way for our purposes. 
 }
\end{figure}

Pure trap {\it weakening} corresponds to $V_f<V_i$, while $L_f=L_i$, and  
pure trap {\it squeezing} to $L_f<L_i$, keeping $V_f=V_i$ (Fig. \ref{potential}). 
We shall next describe the efficiency of atomic Fock state preparation 
by mixed trap reduction ($V_f<V_i$ and $L_f<L_i$) 
keeping constant the  relative smoothness parameter $\tilde{\sigma}_i=\tilde{\sigma}_f$, 
and going from the $U_i$ to the $U_f$ families of traps. This procedure 
allows us to apply squeezing of the potential up to any desired value of $L_f$ keeping its bathtub shape. 

Figure \ref{figFockdistribution} illustrates the full counting statistics of the resulting state in different limits of a trap reduction scheme. 
Both pure weakening and squeezing fail to produce an atom-number state since the condition in Eq. (\ref{cond})  is not fulfilled.
It was shown in \cite{DM08} that this limitation arises as the result of truncation of the final state both in coordinate (pure weakening) or momentum (pure squeezing) space, with respect to the  desirable Fock state $|N=C_f\ra$.
However, this state, whose  full-distribution reduces to a Kronecker delta $\delta_{n,C_f}$ (see Eq. (\ref{Fockdistribution})), can be obtained by combining both strategies, as shown in the middle panel. The final trap is then perfectly filled by the pure atom-number state.
In what follows we shall characterize the efficiency of the method just by the mean and atom number variance of the prepared state, see Eqs. (\ref{mean}) and (\ref{variance}).  The normalized variance, $\sigma_{N_f}^2/\la N_f\ra$ allows us to distinguish between the sub-Poissonian, Poissonian, and super-Poissonian statistics whenever it is lower, equal or greater than $1$, respectively.

\begin{figure}[t]
\includegraphics[width=8.5cm,angle=0]{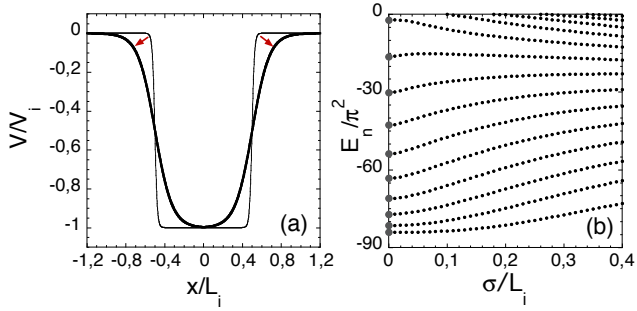}
\caption{\label{specsmooth}
 Effect of the smoothness of the trapping potential (a) on the spectrum (b). As $\sigma$ increases the density of states concentrates near the brim of the trap and more bound states appear. The spectrum in the low panel is obtained for a potential $V=(10\pi)^2$  (in units of $\hbar^2/2mL^2$) .
}
\end{figure}
\begin{figure}[t]
\includegraphics[width=8cm,angle=0]{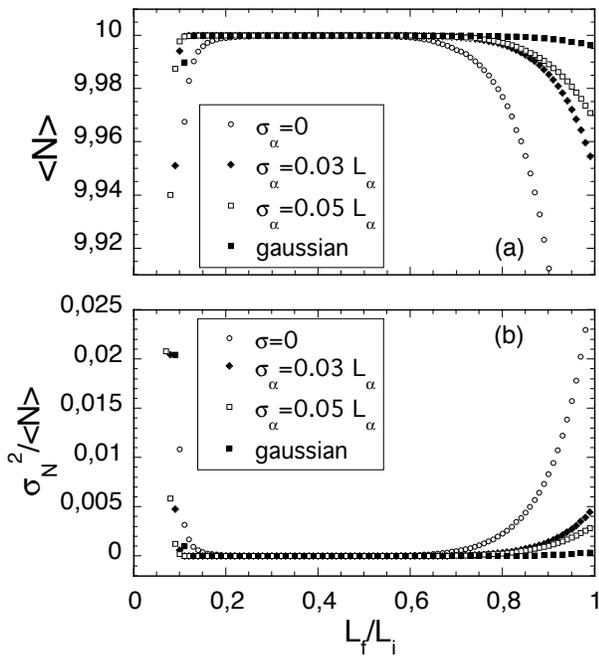}
\caption{\label{smoothness}
 Asymptotic mean value (a) and variance (b) of the atom number distribution of a Tonks-Girardeau gas obtained by sudden weakening-squeezing as a function of the width ratio between the final and initial trap $L_f/L_i$. The process is robust for different values of the smoothing parameter $\sigma$.
The initial state is assumed to be in the ground state. $U_\alpha$ are chosen such that $C_i=100$ and $C_f=10$, this means that for the bathtub potential $U_i=(100\pi)^2$ and $U_f=(10\pi)^2$, while for the Gaussian,  $U_i=(28\pi)^2$ and $U_f=(8\pi)^2$. In all cases we assume $N_i=C_i=100$.
}
\end{figure}
Let us now consider different trap geometries. Generally, the effect of the smoothness of the potential is to increase the density of states near the brims, where the spacing between adjacent energy states is reduced, see Fig. \ref{specsmooth}. As a consequence, a higher control of the depth of the potential would be required. Nonetheless Fig. \ref{smoothness} shows that by increasing the smoothness, 
the Fock state creation condition (\ref{cond}) is actually satisfied for a broader range of parameters which includes conditions nearer pure weakening and pure squeezing. This is because the initial state is spread out along the same region in configuration space as the final one; moreover the looser confinement reduces the momentum components of the final state, which can be resolved more easily by the initial state. 

We might conclude that an invariably efficient strategy for the sudden transition between 
a $U_i$ and $U_f$ trap families, is achieved by reducing to half the width of the 
initial trap and reducing the depth to the trap accordingly, so as to achieve the desired $U_f$ and capacity $C_f$, 
\beqa
L_f\approx\frac{L_i}{2},\qquad V_f\approx\frac{U_f}{U_i}\frac{V_i}{4}, 
\eeqa 
which warrants the preparation of the Fock state $|C_f\ra$ corresponding to the $U_f$-family.   
\section{Non-zero temperature}

%
%
%
\begin{figure}[t]
\includegraphics[width=8cm,angle=0]{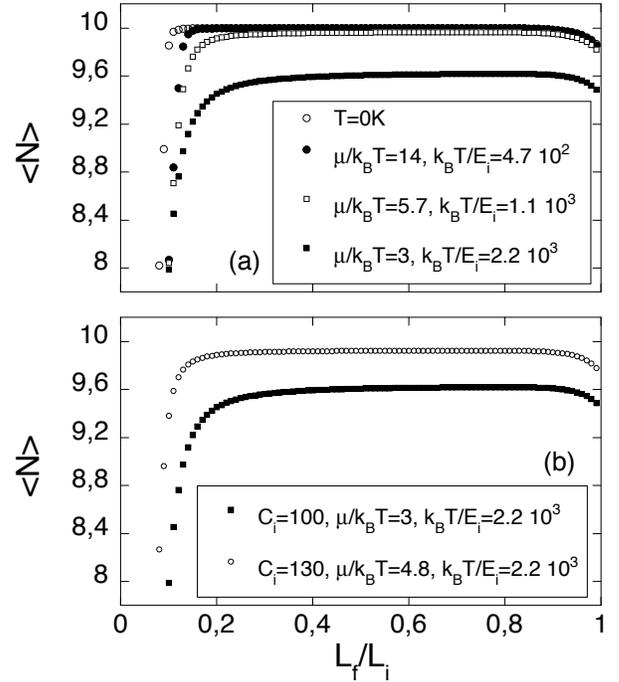}
\caption{\label{pot}
 Effect of temperature on the asymptotic mean value of the atom number distribution of a Tonks-Girardeau gas obtained by sudden weakening-squeezing for the case of a square potential as a function of the width ratio between the initial and final trap. In plot (a) $U_i=(100\pi)^2$ and $U_f=(10\pi)^2$  so the capacities $C_\alpha$ are the same as in Fig. \ref{smoothness}, but the initial occupation is now chosen to be $N_i=0.8C_i$. As we increase the temperature, $\mu /k_BT \le 5$ the method starts to fail, but this can be improved increasing the capacity of the initial well, as shown in plot (b), where the weakening-squeezing process is applied at the same temperature but two different initial traps: $U_i=(100\pi)^2$ used in the previous plot and $U_i=(130\pi)^2$, both with $N_i=0.8C_i$. ($E_i=\hbar^2\pi^2/2m{L_i}^2$.)
 }
\end{figure}
%
%
%
%
%

The above formalism can be generalized in a straightforward way to account for the atom number distribution resulting from an arbitrary initial state at non-zero temperature.
It suffices to redefine 
\beqa
\widehat{\Lambda}_i=\sum_n\pi_n|\varphi_n^i\ra\la\varphi_n^i|
\eeqa
which, in general, is not a projector now,
where $\pi_n$ is the occupation probability of the state $|\varphi_n^i\ra$. In the ground state of the TG gas $\pi_n=1$ $\forall n=1,\dots,N_i$ and $\pi_n=0$ otherwise. For a thermal state, the Fermi-Dirac weights $\pi_n=\{\exp[\beta(E_{n}^{i}-\mu)]+1\}^{-1}$ (with $\beta=1/k_BT$ where $k_B$ is the Boltzmann constant and $T$ the absolute temperature) result due to the effective Pauli exclusion principle mimicked by bosons in the TG regime \cite{BM04}.   Notice the normalization $\sum_n\pi_n=N_i$.
The preparation of a Fock state by a sudden change of the trap will still be feasible as long as $\widehat{\Lambda}_f\subset\widehat{\Lambda}_i$.
The numerical results in Figure \ref{pot} (upper panel) illustrate the degradation of the quality of final state with increasing temperature. However, the lower panel shows that this
negative effect of temperature can be compensated by starting from a ``bigger'' initial trap
with larger capacity.    

\section{Discussion and conclusion} 
We conclude that the controlled preparation of atomic Fock states  
in the strongly interacting (Tonks-Girardeau) regime can be achieved by combining weakening and squeezing of the trapping potential. The process is robust with respect to the smoothness of the 
potential trap, and moreover the deteriorating effect of increasing temperature can be compensated by enlarging the capacity of the initial trap. However, it is still an experimental challenge to get to the strong TG limit which must translate into a correction to the fidelity. By contrast, non-interacting polarized Fermions would be an ideal system for Fock state preparation. For ultracold fermions, due to the wavefunction antisymmetry, s-wave scattering is forbidden and generally p-wave interactions can be neglected so that the gas is non-interacting to a good approximation. Such type of gases can be prepared in the laboratory with linear densities of the order $0.2-2$ $\mu$m$^{-1}$ for which the polarization remains constant 
in a given experiment \cite{Koehl}. For such gases the trap reduction technique can be directly extended to two and three dimensions.


\begin{acknowledgments}
We acknowledge discussions with Martin B. Plenio and J\"urgen Eschner.
This work has been supported by Ministerio de Educaci\'on y Ciencia 
(FIS2006-10268-C03-01, FIS2008-01236) and the Basque Country University (UPV-EHU, GIU07/40).
The work of MGR was supported by the R. A. Welch Foundation and the National Science Foundation.

\end{acknowledgments}


\end{document}